\documentclass[epj,referee]{svjour}

\usepackage{graphics}

\begin{document}

\title{A New Method to Compute the Configurational Entropy in Glassy
Systems} 

\author{Barbara Coluzzi\inst{1} \and Andrea Crisanti\inst{2}
        \and Enzo Marinari\inst{2,3} \and Felix Ritort\inst{4} 
	\and Andrea Rocco\inst{4}
\thanks{\emph{Present Address: Centrum voor Wiskunde en
              Informatica, Postbus 94079, 1090 GB Amsterdam, 
	      The Netherlands.}}
}

\institute{ \inst{1} {\em Service de Physique Th\'eorique}, CEA -
                      Saclay - Orme des Merisiers, 91191 Gif sur Yvette, 
		      France.\\ 
            \inst{2} Dipartimento di Fisica, Universit\`a di Roma 
		     {\em La Sapienza}, Istituto Nazionale Fisica della 
		     Materia Unit\`a di Roma I and SMC, \\
		     P.le Aldo Moro 2, I-00185 Roma, Italy.\\ 
            \inst{3} Istituto Nazionale Fisica Nucleare, Sezione di Roma 1\\ 
	    \inst{4} Physics Department, Faculty of Physics University of 
	             Barcelona, Diagonal 647, 08028 Barcelona, Spain.  
          } 

\date{Received: date / Revised version: date}

\abstract{
We propose a new method to compute the configurational
entropy of glassy systems as a function of the free energy of valleys
at a given temperature, in the framework of the Stillinger and Weber
approach.  In this method, which we call free-energy inherent
structures (FEIS) approach, valleys are represented by inherent
structures that are statistically grouped according to their
free-energy rather than the energy as is commonly done in the standard
procedure.  The FEIS method provides a further step toward a
description of the relaxational behavior of glassy systems in terms of
a free energy measure. It can be used to determine the character of
the glass transition as well as the mode coupling and the Kauzmann
temperatures. We illustrate the usefulness of the method by applying it to
simple models of glasses and spin glasses.
}

\PACS{ {64.70.Pf}{Glass
transitions} \and {75.10.Nr}{Spin-glass and other random models} \and
{61.20.Gy}{Theory and models of liquid structure} } 
% end of PACS codes } 
%end of abstract
%
\maketitle
\section{Introduction}
\label{sec:intro}
\noindent
Spin glasses and structural glasses are characterized by the presence
of a complex structure of stable and meta-stable states \cite{BOOK}.
In particular, the free energy landscape is characterized by a large
number of valleys which in spin-glass theory are also commonly referred to
as basins, pure states, phases or ergodic components. These
valleys contain a large number of configurations (typically growing
exponentially with the size of the system) with macroscopic properties
(for example the energy or the magnetization) that in general depend
on the specific valley.  
The complexity of the configurational space topology 
reflects itself into a complex dynamical behavior commonly 
referred to as {\it glassy behavior}.  These ideas trace back to more that $30$ 
years ago to a seminal paper of Goldstein \cite{Gold}, 
who suggested that 
the dynamics of a supercooled liquid can be understood in  terms of a 
diffusive 
process between different valleys. At low temperature the dynamics slows down 
since the system gets trapped for long time in a valley
and the slow long time relaxation is only governed by the 
inter-valley motion.

Along the same lines, but in a different context, Stillinger and Weber
(hereafter referred as SW) proposed in the early eighties to identify
each valley with its minimum, called Inherent Structure (IS) and to
build up an IS-based analysis \cite{IS}. Each valley contains all
those configurations that map into the IS by a steepest descent
dynamics.  On the one hand,  such a coarse-grained description of the
phase space may be useful to grasp the universal and generic features
of glassy dynamics. On the other hand, replacing each valley with a
single configuration implies of course a strong reduction of
information. A relevant question then is to understand under which
conditions the resulting IS-based dynamics is equally well
representative of the off equilibrium behavior of the system. From
purely theoretical grounds, the decomposition of the phase space into
non-overlapping valleys is always possible. However, its operative definition
may be a rather subtle issue.  In mean-field theory, where ergodicity
breaking holds, the decomposition of the phase space into valleys has
a clear physical meaning. In that case one can define as
configurations belonging to the same valley those such that starting
from any of them, any other configuration in the same valley can be
reached (or get close as we like for continuous systems) in a finite
time.  In the case of short ranged systems, however, the extension of
this definition remains quite unclear. In fact, in the case of finite
dimensional systems, even in presence of Replica Symmetry Breaking
(RSB), activated processes must be included in the physical picture.

An alternative point of view rests on a dynamical definition of the
concept of valley \cite{SDYN}.  Metastable states or valleys ${\cal
V}$ can be defined by introducing the observation time $\tau$ and
looking at the cumulative probability $P^{\cal V}(\tau)$ that the
system starting in any configuration contained in the valley
${\cal V}$ at time $t=0$ escapes from ${\cal V}$ at or before time
$\tau$. The valley is well defined \cite{PA} if there exists a
reasonably small significance level $p$ such that the following
inequality is satisfied: 
\begin{equation}
P^{\cal V}(\tau) \leq p,
\end{equation}
The escape probability, and hence the partition, in general depends
on the external parameters such as fields or temperature and on the
system size $N$. If for a valley it happens that 
\begin{equation}
\lim_{\tau \rightarrow \infty} \lim_{N
\rightarrow \infty} P^{\cal V}(\tau)=0
\end{equation}
then the valley is absolutely confined \cite{PA} in the thermodynamic
limit and corresponds to an ergodic component.
This is the typical case for mean-field models since free-energy barriers
between valleys growth with the system size.

For non-mean-field systems valleys are not, in general, absolutely confined 
and their identification is not trivial. A possible strategy to deal with 
those systems is that of looking for possible simple limiting cases and 
using them to partition the phase space. 
The SW scheme belongs to this
class. Indeed since at $T=0$ any barrier cannot be surmounted, it is clear
that all valleys are absolutely confined at $T=0$, making their identification
simpler.  The SW scheme identifies  valley with a $T=0$ dynamics, 
therefore we can regard the SW scheme as a $T=0$ partitioning.

In the original SW scheme different valleys are classified according to the 
energy of the minimum (IS). This picture can be somewhat misleading since 
it plays down the role of entropy. In this paper we shall 
discuss the more physical classification in terms of 
{\it Free Energy Inherent Structure} (FEIS). We also introduce 
a direct method, which to our knowledge is new, of computing 
the free energy of valleys. We emphasize that the FEIS classification
is more general than the usual IS-energy one. The two classifications
lead to the same information  only if different valleys have similar 
entropies, which in general may not be the case. 

Strictly speaking, the partitioning depends on temperature, this is
particularly relevant for the dynamical properties of the systems. 
We do not address here this problem since we are interested in
understanding the role of entropy once a partition is given. 
For this reason we merely assume the validity of the SW scheme which has the advantage 
of being quite easy to be implemented.

The paper is organized as follows. In Section \ref{sec:part} we discuss
the phase space partitioning and identify the relevant
physical quantities. The method to compute the FEIS is described in
Section \ref{sec:valley} and is applied to two 
representative models, the Sherrington-Kirkpatrick (SK) model and the 
Random Orthogonal (ROM), in Sections \ref{sec:sim} and \ref{sec:res}.
The choice of these model is motivated by their different 
phase space organization. Finally
Section \ref{sec:con} contains conclusions and some perspectives.

\section{Partitioning the phase space into valleys: the free-energy landscape}
\label{sec:part}

In general, given a decomposition of the phase space
into non-overlapping valleys, 
we can write the
partition function ${\cal Z}$ as
\begin{equation}
{\cal Z}(T)
  = \sum_{\cal C}\,\exp\bigl( -\beta {\cal H}({\cal C})\bigr)
  = \sum_{\cal V}\exp\bigl(-\beta {\cal F}_{\cal V}(T)\bigr)\ ,
\label{eq0}
\end{equation}
where ${\cal C}$ denotes the spin configurations, ${\cal H}$ is the
Hamiltonian and $\beta\equiv \frac{1}{T}$ is the inverse
temperature. ${\cal F}_{\cal V}(T)$ is the free energy of the valley
${\cal V}$,
\begin{equation}
  {\cal F}_{\cal V}(T) \equiv -\frac{1}{\beta}
  \log\bigl(\sum_{{\cal C}\in {\cal V}}\exp(-\beta  {\cal
  H}({\cal C}))\bigr).
\end{equation}
Expectation values of observables can be defined 
in a given valley ${\cal V}$ by restricting the Boltzmann
measure to configurations belonging to that valley
(restricted ensemble \cite{PA}):
\begin{equation}
\langle {\cal A} \rangle_{\cal V}\equiv
\sum_{{\cal C}\in {\cal V}}{\cal A}({\cal
C})\exp(-\beta {\cal H}({\cal C})+\beta {\cal F}_{\cal V}(T)).
\end{equation}
Note that 
\begin{equation}
\langle {\cal A} \rangle = \sum_{\cal V} \langle {\cal A} \rangle_{\cal V}
\exp{\left ( -\beta {\cal F}_{\cal V}(T)+\beta {\cal F}(T) \right)},
\end{equation}
where ${\cal F}$ is the total free energy of the system.
After these definitions it is natural to consider the entropy of the
valley $\cal V$, defined as 
\begin{equation}
S_{\cal V}(T) \equiv \beta(\langle E
\rangle_{\cal V}(T)-{\cal F}_{\cal V}(T)).
\end{equation}
In principle the partitioning of
the phase space into valleys allows to extend all thermodynamic
relations to more general valley-dependent ones. As stated above, the
relevant question is whether a given partitioning of phase space has a
physical meaning.

Recent results about the dynamics of spin glasses have shown that a
modified form of the fluctuation-dissipation theorem (hereafter
referred to as FDT) is satisfied in the aging regime~\cite{FDT}. For
structural glasses or one-step RSB spin glasses these violations are
asymptotically described by a single time-scale which can be related
to an effective temperature. Violations of FDT are tightly related to
the spectrum of off-equilibrium fluctuations and the dependence on a
single effective temperature has been advocated as the consequence of
the cooperative character of the dynamics \cite{CR1}. Although FDT
violations have been computed in detail only in mean-field systems
there is evidence that they yield a physically valid description of
the free energy fluctuations in the aging state also for short-ranged
systems \cite{SR}. The fact that FDT violations can be well
rationalized in a mean-field framework \cite{FV} (the effective
temperature is related to the function $x(q)$ in the Parisi solution
of spin glasses) suggests that some features of the mean-field theory
can be extended to realistic systems \cite{CR1}. A key concept
describing this mean field behavior is the {\em configurational
entropy} (also called {\em complexity}) ${\cal S}_c(f,T)$ defined as
the number of valleys with a given free energy per site $f$.  The
configurational entropy counts the number of valleys accessible to the
system in the same way as the usual entropy in Boltzmann theory counts
the number of configurations with a given energy. Therefore, the
information that it contains has physical meaning if an
equiprobability hypothesis (equivalent to that of equilibrium ensemble
theory) holds for spin glasses and glasses in the off-equilibrium
regime \cite{CR1}. This hypothesis assumes that valleys with a given
free energy $f$ have the same weight and can be treated on the same
footing as far as the dynamical properties of the system are
concerned.  The implications of such an assumption for the description
of complex landscapes in glassy systems are very strong. It provides a
flat measure for the dynamics of glassy systems similar to the Edwards
measure for granular media \cite{EDW,BarKurLorSel02}.

In this framework a description of the off-equilibrium dynamics in
terms of the complexity has been successfully obtained for structural
glass models \cite{ST} within the IS formalism \cite{IS} where each
valley is identified by its minimum IS.  The physical assumption
behind the validity and usefulness of this coarse grained description
(where configurations are grouped into different IS) is that the
time-scale of inter-valley motion is much larger than the time-scale
of intra-valley motion. If typical valleys are locally similar, i.e.,
they can have different energies but similar entropies, then one may
simply count them from the probability distribution of the IS energies
and obtain the complexity, apart from an unknown constant.  This last
assumption can be relaxed by estimating the entropies of the valleys
by treating fluctuations with the harmonic approximation
\cite{CPVM,SKT,BUHE,PO,SA,MO}, or by more realistic approximations
which take anharmonicity into account (unfortunately this procedure is not
applicable to discrete systems).  It should be noted that that
such a method for evaluating the complexity in structural glasses
stems out naturally from a recently proposed thermodynamical approach
\cite{MP,MARC,CPVM} to the liquid glass transition, within the
approximation that free energy valleys are well described by the
corresponding IS valleys.

We note that the utility of the SW decomposition is less evident in
the case of Full Replica Symmetry Breaking (FRSB) spin glasses where
the structure of valleys is more sensitive to temperature and no clear
time scale separation exist, making the application of a $T=0$
decomposition to finite $T$ questionable.  Nevertheless a recent
numerical study \cite{CMPR} shows that the coarse graining from
configurations to IS preserves the physical properties of the system.
We shall see that a statistical description of the free-energy
landscape made in terms of the free energy of valleys (rather than their energy)
is still useful also for 
FRSB spin glasses at finite temperature, and that
the corresponding complexity ${\cal S}_c(f,T)$ gives a proper description of the 
glassy phase.

In the next Section we shall propose a method to numerically compute
the valley free energy suitable for discrete systems.  This method
will provide us with the explicit dependence of the configurational
entropy on both temperature and free energy.  Moreover it directly
gives $s_c$ (including also undetermined global constants) without
using any approximation for the intra-valley entropy (such as the
harmonic approximation discussed in the last but one paragraph),
improving the results obtained with other
methods~\cite{CPVM,SKT,BUHE,PO,SA,MO,CRIS}.  To our knowledge this is
the first time that a method to compute the statistical
distribution of the free energies of the IS valleys has been proposed
and tested.

\section{Description of the free-energy inherent structures (FEIS) method}
\label{sec:valley}
\noindent
We consider the IS decomposition introduced by Stillinger and Weber (SW) \cite{IS}, where each valley ${\cal V}$ is labeled by the
corresponding minimum (IS), and denote its free energy per site by
\begin{equation}
f_{\rm IS}\equiv
\frac{1}{N} {\cal F}_{\cal V}.
\end{equation}
Hereafter small letters will denote
intensive quantities. 

Our goal is to compute the complexity ${\cal S}_c(f,T)$ 
without making any assumption on the similarity of  valleys.  
In the context of Lennard-Jones glasses, Speedy \cite{SP}
has proposed to compute the free energy of a valley by adding to the
Hamiltonian a coupling term between the configuration that labels the 
IS and the run-time configuration, $-\epsilon q$, $q$ being
an appropriately defined overlap.
A similar approach
in which the complexity is evaluated by coupling the run-time configuration
to a generic configuration of the valley was considered in the first
reference in \cite{CPVM}. By taking the
thermodynamic limit first and extrapolating later to $\epsilon\to 0$ it
is possible to restrict the measure to those configurations ${\cal C}$
belonging to the IS valley. This procedure is laborious, since it requires
extensive simulations for different values of the coupling $\epsilon$
and for each different IS.  

Here we propose a different strategy based on the probabilistic 
definition \cite{MA} of the valley free energy. The dynamical evolution 
of the system defines a probability measure $p_{\cal V}$ over the
valleys.
In the case of an ergodic dynamics,
and assuming that the observation time is 
$\tau_{tot} \gg \tau_{eq}$,
the statistical weight of the single valley is obtainable directly
from (\ref{eq0}) and the Boltzmann {\it a priori} equiprobability hypothesis
\cite{MA}:
\begin{equation}
\label{pval}
{p}_{\cal V}(T)=\frac{\tau_{\cal V}}{\tau_{tot}}= 
\exp{\left (-\beta {\cal F}_{\cal V}(T)+\beta {\cal F}_{\rm eq}(T) \right)},
\end{equation}
where $\tau_{\cal V}$ denote the time spent by 
system in the valley ${\cal V}$ during the total observation time
$\tau_{tot}$, and ${\cal F}_{\rm eq}(T)$ the equilibrium free energy.
Alternatively eq. (\ref{pval}) can be derived from the 
information theory as the ensemble which minimizes the information
to specify one valley when we only know that the
system is in contact with a heat bath at $T$ 
({\it unbiased component ensemble}) 
\cite{PA}.
In either case the probability to find at temperature $T$ an IS with free
energy density equal to $f$ is given by
\begin{eqnarray} 
\nonumber
{\cal P}(f,T) &=& \langle\delta(f-f_{\rm IS})\rangle=
\sum_{\rm IS}\delta\left(f-f_{\rm IS}\right) \ p_{\rm IS}\left(T\right)
\\
              &=& \frac{g(f,T)\exp\left(-\beta N f\right)}{\sum_{\rm IS}
               \exp\left(-\beta N f_{\rm IS}\right)}\ ,
\label{eq3}
\end{eqnarray}
where $g(f,T)$ is the density of IS with free energy density $f$
which defines the
extensive complexity 
\begin{equation}
\label{eq3a}
{\cal S}_c(f,T)\equiv \log( g(f,T)),
\end{equation}
and $p_{\rm IS}(T)$ is given by (\ref{pval}).
The denominator in (\ref{eq3}) is the equilibrium partition function
\begin{equation}
\label{eq3bb}
{\cal Z}(T)=\exp(-\beta N f_{\rm eq}(T)).
\label{eqZ}\end{equation}
Equations (\ref{eq3})-(\ref{eq3bb}) lead to the relation
\begin{equation}
  s_c(f,T) =
  \frac{\log\bigl[{\cal P}(f,T)\bigr]}{N} +\beta(f-f_{\rm eq}(T))\ .
\label{eq4}
\end{equation}
The key formula in this expression is the probability ${\cal P}(f,T)$
which can be estimated by computing $f_{IS}$ from (\ref{pval}) with the
following procedure.  After having thermalized the system at temperature
$T$ for $t_{\rm therm}$ Monte Carlo (MC) steps, a steepest descent
procedure is used every $t_{\rm run}$ MC steps for $N_{\rm run}$ to
identify the IS \cite{FOOT1}.  
The interval $t_{\rm run}$ is chosen to be of the same order as
$t_{\rm therm}$ to ensure that configurations obtained at the end of
each period are uncorrelated.

If the number of different IS is not too large and 
$N_{\rm run}$ is taken large enough to ensure
that each IS has been visited a substantial number of times, 
then $p_{\rm IS}$ can be estimated as
\begin{equation}
\label{pis}
p_{\rm IS}(T)= \frac{N_{\rm IS}}{N_{\rm run}}
\end{equation}
where $N_{\rm IS}$ is the number of times the given IS has been found
($\sum_{\rm IS}N_{\rm IS}=N_{\rm run}$).  
From (\ref{pval}) the IS
free energy now reads
\begin{equation}
f_{\rm IS}(T)= -\frac{T}{N}\log\Bigl (\frac{N_{\rm IS}}{N_{\rm run}}\Bigr) +
                   f_{\rm eq}(T).
\label{eq5}
\end{equation}
The equilibrium free energy $f_{\rm eq}(T)$ can be computed 
by performing a different MC run and integrating the energy of the system from infinite 
temperature limit down to the working temperature $T$:
\begin{equation}
\beta f_{\rm eq}(T)= \int_{0}^{\beta} d\beta' e_{\rm eq}(\beta') - 
s_{\rm eq}(\beta=0).
\end{equation}
Finally from the 
value of $f_{\rm IS}$ it is easy to construct the histogram ${\cal P}(f,T)$
and using eq. (\ref{eq4}) compute $s_c(f,T)$.  

This approach has an important difference with the usual SW
decomposition. In the standard SW method IS with 
the same energy $e_{IS}$ are assumed to occur with the same
probability. However, in the FEIS method IS with the same free energy
$f_{IS}$ are grouped together. Compared to (\ref{eq3}), in the standard SW procedure we have:
\begin{eqnarray} 
\nonumber
{\cal P}(e,T) &=& \langle\delta(e-e_{\rm IS})\rangle=
\sum_{\rm IS}\delta\left(e-e_{\rm IS}\right) \ p_{\rm IS}\left(T\right)
\\
              &=& \frac{g_e(e)\exp\left(-\beta N f(e,T)\right)}{\sum_{\rm IS}
               \exp\left(-\beta N f_{\rm IS}(e,T)\right)}\,
\label{eq3b}
\end{eqnarray}
and therefore 
\begin{equation}
  s_c(e) =
  \frac{\log\bigl[{\cal P}(e,T)\bigr]}{N} +\beta(f(e,T)-f_{\rm eq}(T))\,
\label{sce}
\end{equation}
with
\begin{equation}
s_c(e)=\lim_{T \rightarrow 0} s_c(f,T).
\end{equation}
For models with continuous degrees of freedom
$s_c(e)$ can be obtained directly from (\ref{sce}) by evaluating 
$f(e,T)$ with, e.g., the harmonic approximation.

For discrete models this decomposition is useful 
whenever the free energy of valleys
with IS energy $e$ has a trivial
dependence on $e$:
\begin{equation}
f(e,T) \simeq e + f_0(T),
\label{apprf}
\end{equation}
with $f_0(T)$ only function of temperature. Note that in this case
one has
\begin{equation}
s_c(f,T)=s_c(e+f_0(T)).
\label{par}
\end{equation}

This approach relies on the hypothesis that all valleys
with the same IS energy have the same relevance to the statistical
properties of the system, so that the frequency of visiting an IS
with a given energy only depends on the total number of IS with 
that energy.
If the approximation (\ref{apprf}) is valid, $s_c(e)$ can be evaluated,
apart from the unknown constant $f_0(T)$, directly
from (\ref{sce}). As the unknown constant in $s_c(e)$ only depends on $T$ (and
not on the energy) one can check the
correctness of the results from the superposition of the appropriately 
shifted curves obtained at different temperatures. 
Usually it is reasonable to
expect that this happens in at least two different situations. First, when the
temperature is low enough that only configurations near the bottom of
the IS valleys contribute; second, when the IS-valleys are narrow
as in REM-like models \cite{DE}.
As we are going to discuss in detail, this is not the case in FRSB spin 
glasses, where the size of valleys must also be taken into account. In such
case the configurational entropy is meaningful only when 
expressed in terms of the free energies of the valleys.

We conclude this Section by noting that the partition function (\ref{eqZ}) can be
written as
\begin{eqnarray}
{\cal Z}(T)& =& 
\sum_{IS} \exp(-\beta N f_{\rm IS}) =\int df \: g(f,T) \exp(-\beta N f)
\nonumber \\
& = & \int df \: \exp(-\beta N \Phi(f,T)),
\label{eqZ2}\end{eqnarray}
with the generalized free energy
\begin{equation}
\Phi(f,T) \equiv f -T s_c(f,T).
\label{phia}
\end{equation}
In the large $N$ limit, the equilibrium free energy is given by
the minimum of $\Phi$:
\begin{equation}
f_{\rm eq}(T)=\min_{f} \Phi(f,T)=f^*(T)-T \: s_c(f^*(T),T),
\label{phib}
\end{equation}
where we used (\ref{eqZ}) and (\ref{eqZ2}).
Under the assumption that the typical IS energy $e^*(T)$ is similar to the equilibrium energy  $e_{\rm eq}(T)$ then the complexity $s_c(f^*(T),T)$ is 
simply given by the
difference between the entropy of the system at equilibrium $s_{\rm
eq}(T)$ and the typical IS entropy $s^*(T)$ (see Ref. \cite{MP,MARC}).  
This allows the numerical computation of
$s_c(f^*(T),T)$ in systems with continuous degree of freedom such as
structural glass models where $s^*(T)$ can be evaluated as the average
entropy of a disordered harmonic solid (or also within more realistic
approximations).

The formalism developed in~\cite{MP,MARC} is appropriate for glassy
systems with a clearcut time-scale separation between the inter-valley
and the intra-valley motion, so that the valley free energy is a well
defined quantity.  Our approach does not suffer from this limitation and
we shall show that $\Phi$ in (\ref{phia}) is the relevant quantity to
look at also in FRSB model, where no clear time-scale separation exists
at finite temperature.  Moreover it is just the behavior of $\Phi$ at
different temperatures which gives information on the nature of the
glass transition allowing to predict the values of the different
relevant temperatures characterizing the glass transition (e.g. the MCT transition temperature or the Kauzmann temperature).

In what follows, we will
show the powerfulness of our method 
using simple spin models of glassy systems.  

\section{Models and Simulation Details}
\label{sec:sim}
\noindent
We have considered two spin glass models: the Sherrington-Kirkpatrick
model \cite{SK} (SK) and the random orthogonal model
(ROM) \cite{ROM}. They are both described by the Hamiltonian
\begin{equation}
{\cal H}=-\sum_{1\le i<j\le N}J_{ij}\,\sigma_i\sigma_j,
\end{equation}
where the spins are Ising spins ($\sigma_i=\pm 1$) and the $J_{ij}$ 
are symmetric ($J_{ij}=J_{ji}$)
quenched random couplings. In the SK model the $J_{ij}$ are
uncorrelated variables with zero mean and $1/N$ variance, while in the
ROM $J_{ij}$ are the elements of a random orthogonal matrix, with $J_{ii}=0$ and
\begin{equation}
\sum_k \, J_{ik}
J_{jk} = 4\delta_{ij},
\end{equation}
We have considered both the SK model with binary couplings $J_{ij}=\pm
1/\sqrt{N}$ and with Gaussian couplings
$P(J_{ij})=(2\pi/N)^{-1/2}\exp(-NJ_{ij}^2/2)$, obtaining very similar results.

The reason for this choice of the SK and ROM is based on the fact that
they describe the two most relevant scenarios of mean-field glassy
systems: the SK model gives the mean-field theory for disordered and
frustrated magnets with FRSB, while the ROM describes structural glasses
in the mode coupling approximation \cite{BOOK}.  Recent work
\cite{CRIS} suggests that the study of finite-sized mean-field systems is a
useful route to investigate activated processes in real systems.  A
statistical analysis of the IS in the SK model has been presented in
\cite{CRIS,CMPR}. 
In particular, it was shown in \cite{CMPR} that the
probability distribution of the overlap $P(q)$, i.e. the order parameter
that describes the FRSB transition, can be computed between IS instead
of equilibrium configurations (weighting the IS with their probability
of occurrence in the simulation $p_{IS}(T)$, which is the weight of the 
corresponding basin at the considered temperature). 
In this sense, the coarse-graining from configurations to IS seems to preserve
the physical properties of the system.
  For the ROM it has been found in \cite{CRIS} that valleys
are statistically identical as they have a very small intra-valley
entropy, and the usual SW decomposition in terms of the energy of the
different IS provides a fairly good statistical description of the
potential energy landscape and the relaxational dynamics.  In the SK
model (and, by extension, all models with continuous RSB), 
contrarily to the ROM, valleys can be very different, so the IS free
energy has important contributions coming from intra-valley fluctuations.

We have considered small volumes, typically $N=32$ and $64$, to avoid 
having too many valleys. In fact the number of stationary points of the
energy surface ${\cal N}_{\rm IS}$ is known to grow exponentially with the 
system size:
\begin{equation}
{\cal N}_{\rm IS}\sim \exp(\alpha N)
\label{nis}
\end{equation}
with $\alpha\simeq 0.2$ for the SK model \cite{BM} and $0.3$ for the ROM
\cite{PARPOT}.  
Because of that, if we want to satisfy the condition
${\cal N}_{\rm IS} / N_{\rm run} \sim {\cal O}(1)$ in (\ref{pis}), $N$
cannot be too large, otherwise ${\cal N}_{\rm IS} / N_{\rm run}=0$ for
those IS inefficiently sampled.  In the case of the SK model for $N=32$
we have ${\cal N}_{\rm IS}\simeq 10^3$ while for $N=64$ typically we
have ${\cal N}_{\rm IS}\simeq 10^5$.  In all cases we used $N_{\rm run}
= 10^6$.  In general, the study of larger sizes requires a very large
amount of memory and CPU time though it should be possible, as 
we will discuss in more detail in the following, to restrict the analysis
to a smaller range of free energy values.

We have also checked that, as expected \cite{FOOT1}, these results do not 
depend on the details of the
algorithm used to determine the IS: we have verified that the
greedy algorithm (which follows the steepest path in the phase space)
and a $T=0$ MC dynamics produce the same results. 
For the finite $T$ simulations we have used a MC update with a simple
Glauber dynamical rule. To reduce sample-to-sample fluctuations for the SK 
model data have been averaged over $10$ samples. 
For the ROM we find very small sample-to-sample fluctuations.
Already for systems of size $N=32$ different samples almost give the same 
results. This is most probably due to the high degree of correlation
among the couplings imposed by the orthogonality requirement.
For this reason, and because simulations turn out to be very long, the data for ROM reported in the figures  
have been averaged over only two samples.

\section{Results and Discussion}
\label{sec:res}
\begin{figure}[tbp]
\begin{center}
\resizebox{0.9\columnwidth}{!}{
\includegraphics{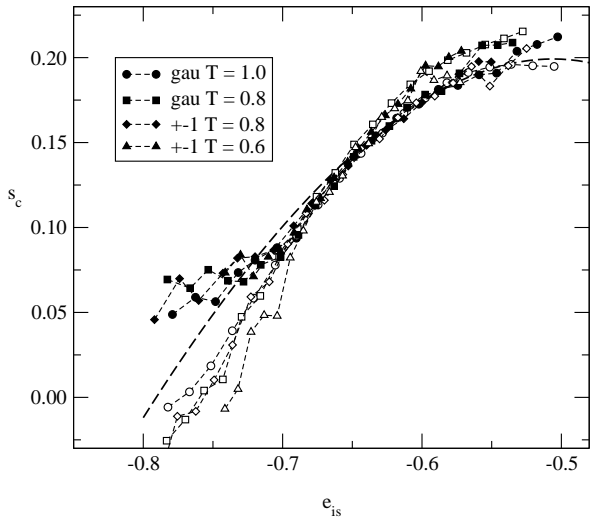}}
\caption{The complexity as a function of the IS energy for the SK
model with $N=64$ at temperatures $T=1.0,0.8,0.6$, for Gaussian and $\pm 1$
couplings. The filled symbols are quenched averages, the empty ones
correspond to annealed averages. The dashed line is the analytic annealed
result \cite{BM}.}
\label{fig1}
\end{center}
\end{figure}

\noindent
In Figure~\ref{fig1} we show the complexity $s_c(e)$ for the SK model as a
function of the IS energy as defined in the usual SW decomposition 
computed from eq. (\ref{sce}).  
We have considered two different
averages: 
\begin{itemize}
\item the annealed average, where we first average ${\cal
P}(e,T)$ over the disorder (average denoted by $\overline{(...)}$), and
the configurational entropy per spin is estimated as 
\begin{equation}
s_c(e) \equiv {\log\bigl( \overline{ {\cal P} (e,T)} \bigr) \over N} 
+ \beta e + const,
\end{equation}
\item the quenched average, where 
\begin{equation}
s_c(e) \equiv {\overline{\log({\cal P} (e,T))} \over N} + \beta e + const.  
\end{equation}
\end{itemize}

In either case we have determined the normalization
constant by superimposing the data to the Bray and Moore \cite{BM}
analytical result, obtained within the annealed approximation. A reasonable
estimate of the size of the error induced from sample to sample
fluctuations is given from the dispersion of the curves at similar
values of the temperature.

The annealed average corresponds to what is known as white average in the
calculation of the number of TAP solutions. For the SK model, a
zero-temperature calculation shows \cite{BM} that the white average 
gives incorrect results below a critical energy $e_c=-0.672$, where in
fact the quenched average is needed. This can be well appreciated in
figure~\ref{fig1}. Two results emerge from this figure: 
\begin{itemize}
\item the two types
of average agree above the predicted value for $e_c$, and clearly
differ below. The strong difference found between
the two kind of average reflects the fact \cite{BM} that the IS minima are
uncorrelated only for $e \ge e_c$ whereas below the breaking of replica 
symmetry should be taken into account.
\item above $e_c$ it is impossible to obtain a good collapse of the
data, and they do not fit the theoretical prediction \cite{BM}.  This
effect, which is absent in the ROM \cite{CRIS} where all data collapse
pretty well in a single curve, suggests that a complexity defined in
terms of the usual SW decomposition does not give a
good description of the free energy landscape of the SK model.
\end{itemize}

To compute the complexity as function of IS free energy
we have estimated the IS free energy from
(\ref{eq5}) and evaluated the complexity using
(\ref{eq4}). 
We note that the finite number of searches introduces 
some ambiguity in the normalization constant. Indeed
since  $N_{\rm run}$ is finite
the IS with probability 
\begin{equation}
p_{\rm IS}(T) < p^0_{\rm IS}=
\exp(-\beta \: N (f^0_{\rm IS} -f_{\rm eq}(T)) = \frac{1}{N_{\rm run}},
\end{equation}
i.e., with free energy $f_{\rm IS}> f^0_{\rm IS}$, are never found. 
Furthermore,  one can assume that $p_{\rm IS}(T)$ is correctly evaluated only 
for IS that have been found at least few times (say $5$ times).
Consequently
$P(f,T)$ is only known for $f < f^0_{\rm IS}$
and this introduces an unknown constant in $s_c$. 
To eliminate this ambiguity we have fixed the
unknown constant for each sample by the use of  (\ref{phib}).
i.e. by imposing
$f_{eq}(T) = \min_f \Phi(f,T)$.  
In all cases the correction to $s_c(f,T)$ 
obtained directly from (\ref{eq4}) was found very small, 
giving us further confidence in
the quality of our sampling. 
Finally we note that in this case
no remarkable difference between the annealed and the quenched average has 
been observed.

\begin{figure}[tbp]
\begin{center}
\resizebox{0.9\columnwidth}{!}{
\includegraphics{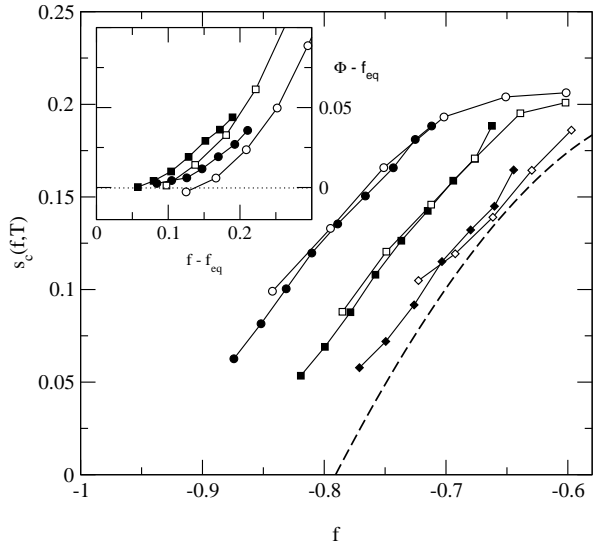}}
\caption{Complexity for the SK model as function of the free energy
for T=1.0 (circles), 0.8 (squares), 0.6 (diamonds). Empty symbols
correspond to $N=32$, filled symbols to $N=64$.  The dashed line is
the analytic result. \cite{BM} In the inset we plot the potential
$\Phi(f,T)-f_{eq}(T)$ as function of $f - f_{eq}$.}
\label{fig2}
\end{center}
\end{figure} 

\begin{figure}[tbp]
\begin{center}
\resizebox{0.9\columnwidth}{!}{
\includegraphics{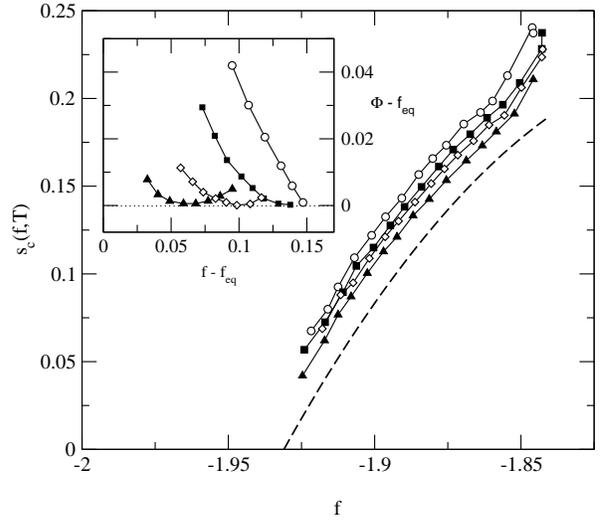}}
\caption{Same as in figure \ref{fig2} for the ROM.
Here $T=0.8$ (circles), 0.7 (squares), 0.6
(diamonds), 0.5 (triangles), and $N=64$. 
The dashed line is the analytic result \cite{PARPOT}.}
\label{fig3}
\end{center}
\end{figure} 

In Figures~\ref{fig2} and \ref{fig3} we show the complexity $s_c(f,T)$ as a
function of the free energy for both the ROM and the SK model, 
 compared
with the corresponding $s_c(e)$, i.e. the $T\rightarrow 0$ limit
computed analytically in Ref. \cite{BM} and \cite{PARPOT}, respectively.  
In the SK model 
the complexity strongly depends on $T$.
On the other hand for the ROM  it is
nearly $T$-independent and, moreover,  
curves at different $T$ are remarkably parallel each other and to the 
zero-temperature limit, in agreement with eq. 
(\ref{par}). This behavior confirms
that for the ROM the intra-valley entropy is very small, giving further
justification to the results obtained in \cite{CRIS} using the
SW decomposition. 

We note that finite-size corrections are small for $s_c(f,T)$ as shown
by the similarity of the results for $N=32$ and $N=64$.  This similarity
is at variance with the strong finite-size effects usually found for
other quantities. 
For instance one finds that finite size corrections
to the $P(q)$ in magnetic field in the SK model are enough strong
to make the two peak structure in the glassy phase hardly visible
for sizes as large as $N=1024$ \cite{BICO} (see also Ref. \cite{FS}). 
However one should note that we are considering
probability distributions of valley dependent quantities so that
finite-size effects directly affect the normalization factor in
$\Phi(f,T)$ (that results in an overall constant shift in $s_c(f,T)$).
This effect is further reduced by fixing the constant in the complexity
self-consistently through eq. (\ref{phib}).  It
seems therefore that the method could be usefully applied also to other
system such as structural glasses, where $N \ge 60$ particles
are known to be enough for a reasonable evaluation of $s_c$ 
(for a detailed
study of finite size effects in a glass forming model see Ref. \cite{BUHE}). 

In each figure the inset shows the
(averaged) shape for the potential $\Phi(f,T)$ at different
temperatures. 
As is clearly seen the shape is different for the two models.
For the SK model $\Phi(f,T)$ has a minimum $f^*(T)$ close to the lower
(measured) 
bound over the free energy support in the whole $T$-range considered. 
The value of $f^*(T)$ converges (for
$N$ large) to $f_{\rm eq}(T)$ corresponding to $s_c(f^*)=0$.  The ROM
shows a different behavior. Here a local minimum appears very close to
the mode-coupling transition temperature $T_{\rm MCT}=0.6$, and moves to
lower free energies as the temperature is lowered \cite{MARC}. The
minimum $f^*(T)$ sticks to $f_{\rm eq}(T)$ at the Kauzmann or static
transition, where $s_c(f^*(T))=0$. From the inset of figure~\ref{fig3}, a
simple linear extrapolation to zero of $f^*(T)-f_{\rm eq}(T)$ as function 
of $T$
gives the estimate $T_K \simeq 0.26$ in excellent agreement with the
theoretical value $0.25$ \cite{PARPOT}. 

\section{Conclusions}
\label{sec:con}
\noindent
The present FEIS approach, i.e. the SW decomposition 
based on inherent
structures but considered in terms of their free energies, do capture
the physics of the system even for finite $T$ and allows for distinguish
between continuous and one step RSB scenarios. It seems particularly
intriguing to us that it describes well also FRSB models, where it does
not correspond to a situation where time-scales are strongly
separated. It moreover gives reasonable evaluations of the relevant dynamic and
static transition temperatures. 

In summary, we have proposed a new method to compute the configurational
entropy as a function of the free energy and temperature. It provides a
way to investigate the free energy landscape of glassy systems which can
be generally applied whenever there is a flat measure describing the
probability to dynamically explore free energy valleys. It can be used
to determine the type of glassy transition (by computing the shape of
the potential $\Phi$), the mode coupling temperature and the Kauzmann
temperatures. 

The main advantage of method is that  it allows to evaluate the complexity
without using any approximation for the intra-valley entropy. Moreover,
one directly obtains an evaluation of the generalized free energy $\Phi$
which contains the physics of the model. On the
other hand, one should note that the method is practically limited 
to relative small size systems. Nevertheless, it is general enough to 
be applied to other
models of structural glasses and spin glasses. It would be particularly
interesting to consider short-range spin-glass models as well as
Lennard-Jones glasses, to achieve a
better understanding of the free energy landscape of generic glassy
systems. 

In particular in structural glasses the total number of IS
is still given by (\ref{nis}) with $\alpha$ depending on the particular
model and on the density, but one finds $0.7 \le \alpha \le 1.2$ for a 
Lennard-Jones model with $N=60$ \cite{BUHE}, which gives a very large 
IS number. Nevertheless it should be still possible to carry out the present 
kind of analysis by fixing an appropriate cutoff on $p_{IS}$, which simply means that 
one is estimating $s_c(f)$ for $f <f^0_{\rm IS}$ instead of on the whole 
free energy range.

As a last remark, it is interesting to note
that at variance with the present result the IS decomposition is not
relevant for describing coarsening models \cite{RO}, since systems
sharing the same $s_c(e)$ display a different dynamical behavior.

\section*{Acknowledgments} B.C. is supported by a Marie Curie fellowship,
contract $n^0$ MCFI 2001-00312. F.R. is supported by the Ministerio de
Educaci\'on y Ciencia in Spain, project BFM2002-3525. A.C and F.R have
been supported by a Italian-Spanish collaboration program (Acciones
Integradas ...)
A. R. is supported
by PAIS 1999 {\it Aging, Slow Dynamics and Glassy Behavior} of INFM
Section G.  We acknowledge useful discussions with G. Parisi.

\end{document}